\documentclass[referee]{raa}            

\usepackage{graphicx,times}             

\begin{document}

   \title{First photometric study of ultrashort-period contact binary 1SWASP J140533.33+114639.1
}

   \volnopage{Vol.0 (200x) No.0, 000--000}      
   \setcounter{page}{1}          

   \author{Bin Zhang
      \inst{1,2,3}
   \and Sheng-bang Qian
      \inst{1,2,3}
   \and R. Michel
      \inst{4}
   \and B, Soonthornthum
      \inst{5}
   \and Li-ying Zhu
      \inst{1,2,3}
   }

   \institute{Yunnan Observatories, Chinese Academy of Sciences (CAS), P. O. Box 110, 650216 Kunming, China; {\it zhangbin@ynao.ac.cn}\\
        \and
             Key Laboratory of the Structure and Evolution of Celestial Objects, Chinese Academy of Sciences, P. O. Box 110, 650216 Kunming, China\\
        \and
             University of Chinese Academy of Sciences, Yuquan Road 19\#, Sijingshang Block, 100049 Beijing, China\\
        \and
             Instituto de Astronom\'{\i}a, UNAM, Ensenada, M$\acute{e}$xico\\
        \and
             National Astronomical Research Institute of Thailand, 191 Siriphanich Bidg. 2nd Fl. Huay Kaew Rd. Suthep District, Muang, Chiang Mai 50200, Thailand\\}

   \date{Received~~2009 month day; accepted~~2009~~month day}

\abstract{In this paper, CCD photometric light curves for the short-period eclipsing binary
1SWASP J140533.33+114639.1 (hereafter J1405) in the $BVR$ bands are presented and analyzed using the 2013 version of the Wilson-Devinney (W-D) code.
It is discovered that the J1405 is a W-subtype shallow contact binary with
a contact degree of $f$ = 7.9$\pm$ 0.5\% and a mass ratio of $q$ = 1.55 $\pm$ 0.02.
In order to explain the asymmetric light curves of the system, a cool star-spot on the more massive component was employed.
This shallow contact eclipsing binary may be formed from a short-period detached system through the orbital shrinkage due to
angular momentum loss.
Based on $(O-C)$ method, the variation of the orbital period was studied using all the available times of the minimum light.
The $(O-C)$ diagram reveals that the period is increasing continuously at a rate of $dP/dt=+2.09\times{10^{-7}}$days yr$^{-1}$,
which can be explained by mass transfer from the less massive component to the more massive one.
\keywords{Stars: binaries: close --
          Stars: binaries: eclipsing --
          Stars: individual (1SWASP J140533.33+114639.1).}
}

   \authorrunning{Bin Zhang et al. }            
   \titlerunning{Contact binary 1SWASP J140533.33+114639.1}  

   \maketitle

%
%

\section{Introduction}
\label{sect:intro}
The W UMa type contact binaries exhibit a sharp period cutoff phenomenon around 0.22 days (Rucinski 1992).
Recent study using the data released by The Large Sky Area Multi-Object Fiber
Spectroscopic Telescope (LAMOST) found that new value is around 0.2 days (Qian et al. 2017).
The ultrashort-period(less than 0.23 days) contact eclipsing binaries (EBs) are important for modern astrophysics:
(1)they can detect the distance of stars relying on their empirical period-luminosity relation (Rucinski 2004);
(2)these binaries offer significant information about origin and evolution of late-type stars
including mass and angular momentum loss, even the merging (Qian et al. 2014; Kjurkchieva et al. 2016).
As the development of technology, more and more contact EBs with short-periods
were discovered by some Sky Surveys in the world (e.g., SDSS, Super-WASP and NSVS).
Till now, some of them have been studied in detail, such as CC Com (Kose et al. 2011; Yang \& Liu 2003), GSC1387-475 (Rucinski \& Pribulla 2008),
1SWASP J015100.23-100524.2 (Qian et al. 2015a), NSVS 4484038 (Zhang et al. 2014) and the
stable red-dwarf contact binary SDSS J001641-000925 (Davenport et al. 2013; Qian et al. 2015b).

1SWASP J140533.33+114639.1 (hereafter J1405) as a short-period contact eclipsing binary (EB) candidate, with a orbital period about 0.225123 days,
was first detected in 2013 (Lohr et al. 2013).
Its LCs present a typical EW-type (nearly equal light minima). Two Micron All Sky Survey (Cutri et al. 2003)
offers its magnitude of V=15.51, J=14.019, H=13.487 and K=13.328, and corresponding color index
are V-K=2.182, J-H = 0.532 and H-K = 0.159 for the system, which imply an average spectral type about K4.
However, there is no spectroscopic element, photometric solution or period research published until now.
In the present paper, the light curves (LCs) are analyzed using the (W-D) program and its photometric solutions are obtained.
All times of minimum light are collected and the period variations are analyzed.
The evolutionary scenario and magnetic activity are also discussed.
\section{Multi-color CCD Photometric Observations}
\label{sect:Obs}
Photometric observations of J1405 were carried
out on 2016 March 17 and 22 using the 84 cm telescope at the Observatorio Astron$\acute{o}$mico Nacional (OAN) at Sierra San Pedro M$\acute{¡äa}$rtir, Mexico.
This relatively small telescope is mainly used for photometric observations, because more than 60\% of the nights at this site have photometric quality (Tapia 2003).
The integration times for each image in $BVR$ bands were 70 s, 40 s and 25 s, respectively.
All the observed images were reduced using the aperture photometric package PHOT of IRAF by Mr. Michel.
Another two stars near the target were chosen as the comparison star and the check star.
The coordinates of the variable star, the comparison star and the check star are listed in Table 1.
Two sets of LCs obtained are plotted in Figure 1.
The LCs are asymmetric and show a weak O'Connel effect (O'Connell 1951), where the
maxima following the primary minima are higher than the other maxima.
And the average observational errors and the amplitudes of the light variation in different bands are listed in Table 2.

\begin{table}
\begin{center}
\caption{Coordinates of J1405, the Comparison and Check Stars}
\begin{tabular}{ccc}
\hline
Stars             &  $\alpha_{j2000}$  &  $\delta_{j2000}$\\
\hline
J1405             & \emph{\emph{$14^{h}05^{m}33^{s}.33$}} &\emph{\emph{$11^{\circ}46^{'}39^{''}.1$}}\\
Comparison        & \emph{\emph{$14^{h}05^{m}38^{s}.23$}} &\emph{\emph{$11^{\circ}41^{'}17^{''}.3$}}\\
Check             & \emph{\emph{$14^{h}05^{m}59^{s}.16$}} &\emph{\emph{$11^{\circ}40^{'}38^{''}.3$}}\\
\hline
\end{tabular}
\end{center}
\end{table}

\begin{table}
\begin{center}
\caption{The Average Observational Errors and the Amplitudes of the Light Variation in March 2016}
\begin{tabular}{c|ccc|ccc}
\hline
Band     &  Date    & Error(.mag) &  $\Delta{m}$(.mag) & Date  & Error(.mag)  & $\Delta{m}$(.mag) \\
\hline
B      &17&0.0071&0.5852&22&0.0102& 0.6646                        \\
V      &17&0.0071&0.5231&22&0.0087& 0.5592                        \\
R      &17&0.0067&0.5480&22&0.0081& 0.5303                        \\
\hline
\end{tabular}
\end{center}
\end{table}

\begin{figure}[!h]
\begin{center}
\includegraphics[width=10cm,height=8cm]{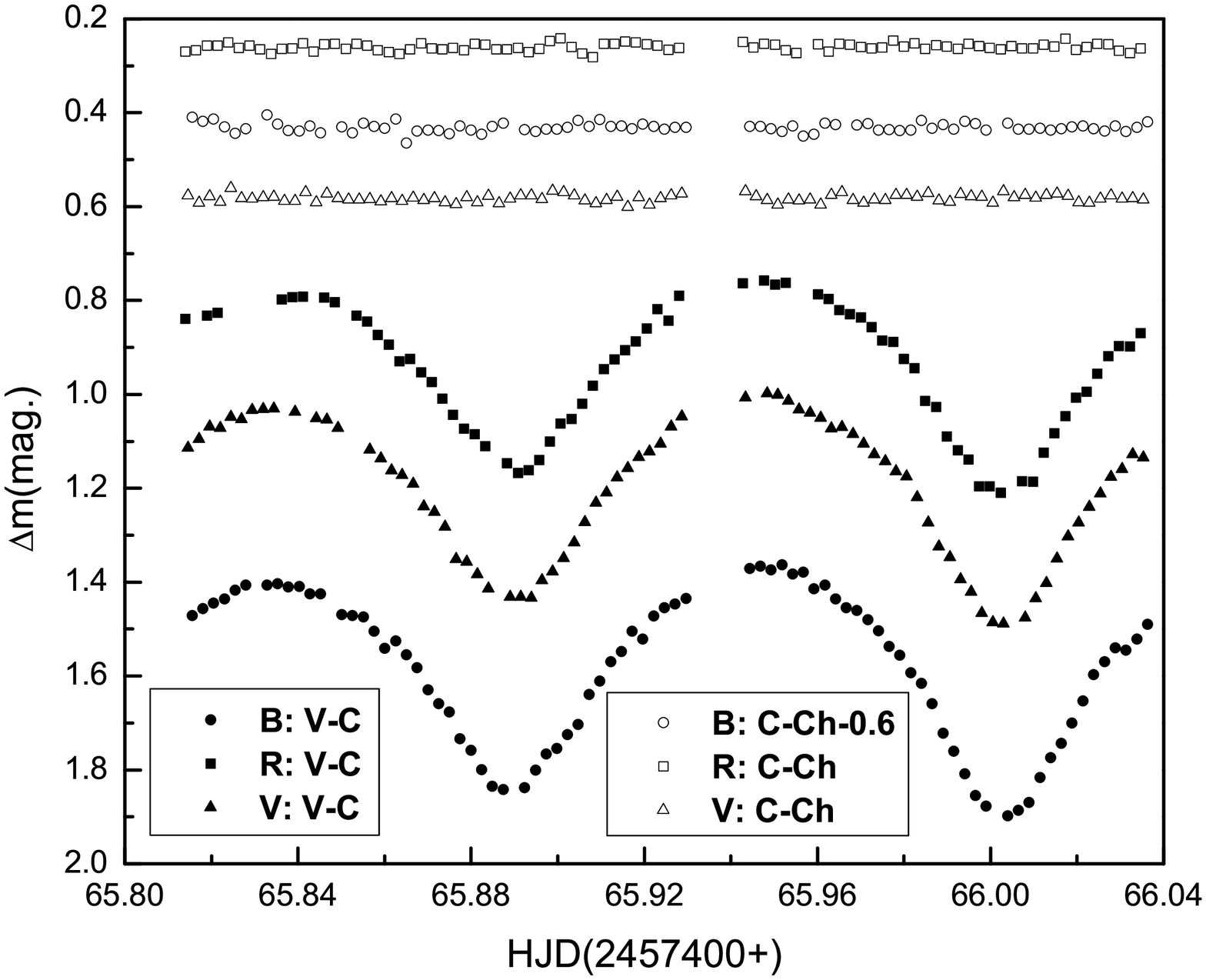}
\includegraphics[width=10cm,height=8cm]{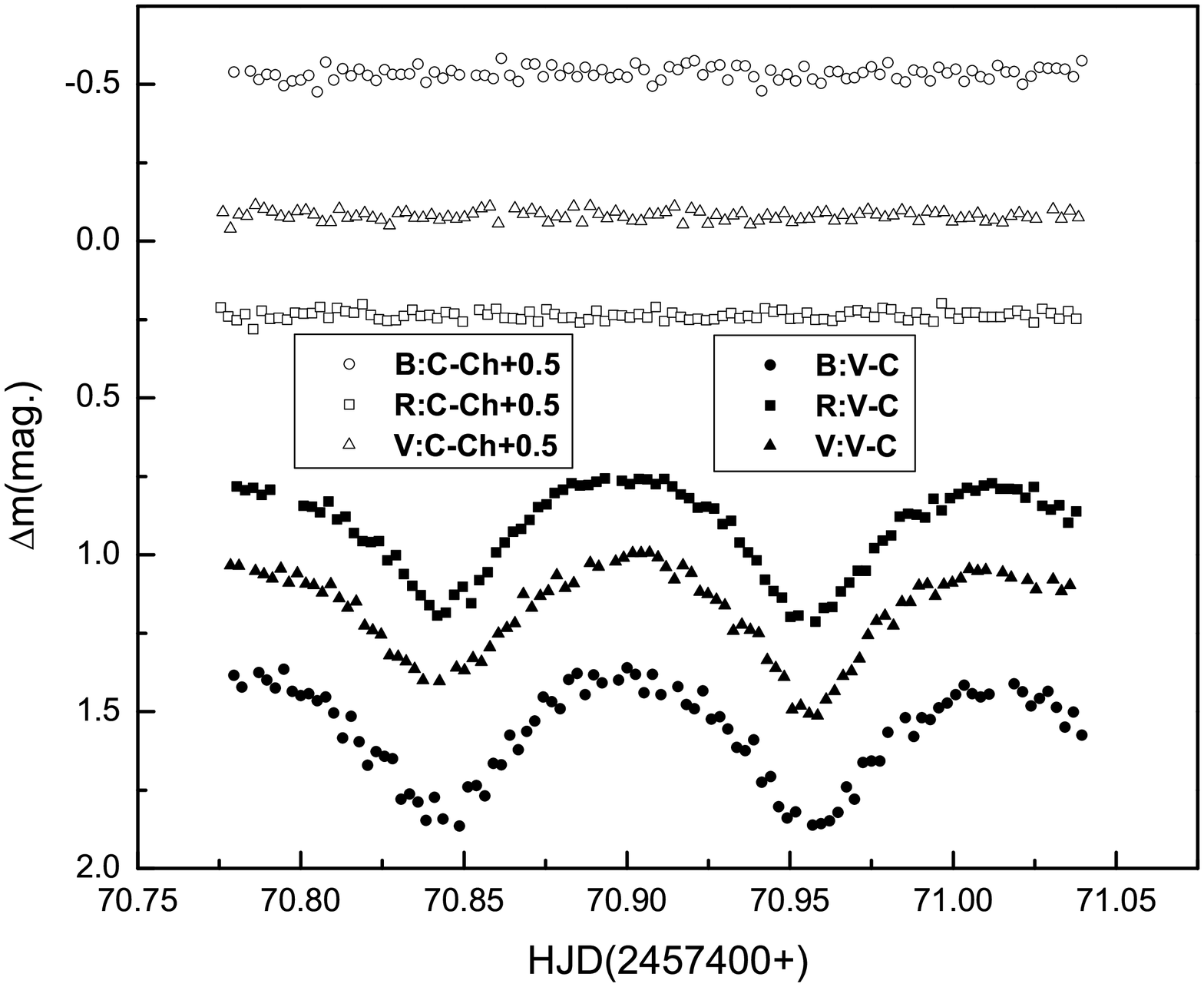}
\caption{The observational LCs of J1405 in $BVR$ bands. The differential light curves of the comparison
star relative to the check star are also plotted.}
\end{center}
\end{figure}

Meanwhile, some new times of light minimum for J1405 were also observed and determined using least-squares parabolic fitting method.
By use of the following linear ephemeris:
\begin{equation}
  Min.(HJD)=2457466.00294(\pm0.00019)+0^{d}.225123\times E.
\end{equation}
the $(O-C)$ values and observational LCs' phase were calculated.
The zeropoint displayed in this linear ephemeris was one of primary eclipse times,
which was determined using the observed data from 84 cm telescope,
and the orbital period we adopted came from Super-WASP EB catalog (Lohr et al. 2013).
All the minima times and the corresponding $(O-C)$ values are listed in Table 3.

\begin{table}
\centering
\tiny
\caption{(O-C) Values of Light Minima for J1405}
\begin{tabular}{lllllllllllllll}\hline\hline
 HJD(2450000+) &(O-C)&E&Filter&Telescope&HJD(2450000+)&(O-C)&E&Filter&Telescope&HJD(2450000+)&(O-C)&E&Filter&Telescope    \\
\hline
7465.8895(02)  &-0.0009 &-0.5  &$BVR$  &84 cm&4941.4512(46) &-0.0224  &-11214 &$V$&INGT  &5279.5906(0.40)&-0.0178  &-9712  &$V$&INGT \\
7466.0029(02)  &0.0000  &0     &$BVR$  &84 cm&4942.5783(57) &-0.0209  &-11209 &$V$&INGT  &5280.4858(0.49)&-0.0230  &-9708  &$V$&INGT \\
7470.8427(08)  &-0.0004 &21.5  &$BVR$  &84 cm&4943.4743(43) &-0.0254  &-11205 &$V$&INGT  &5281.6167(0.65)&-0.0178  &-9703  &$V$&INGT \\
7470.9563(06)  &0.0006  &22    &$BVR$  &84 cm&4944.6026(31) &-0.0228  &-11200 &$V$&INGT  &5290.6089(0.50)&-0.0305  &-9663  &$V$&INGT \\
4450.0010(...) &-0.0291 &-13397&$V$    &INGT &4945.5019(30) &-0.0239  &-11196 &$V$&INGT  &5291.5206(0.44)&-0.0193  &-9659  &$V$&INGT \\
4530.6109(42)  &-0.0132 &-13039&$V$    &INGT &4949.5605(34) &-0.0175  &-11178 &$V$&INGT  &5292.6417(0.42)&-0.0238  &-9654  &$V$&INGT \\
4532.6232(26)  &-0.0271 &-13030&$V$    &INGT &4950.4561(26) &-0.0224  &-11174 &$V$&INGT  &5293.5505(0.36)&-0.0155  &-9650  &$V$&INGT \\
4533.5289(49)  &-0.0218 &-13026&$V$    &INGT &4955.6460(73) &-0.0104  &-11151 &$V$&INGT  &5294.6693(0.24)&-0.0223  &-9645  &$V$&INGT \\
4535.5631(35)  &-0.0138 &-13017&$V$    &INGT &4963.5170(47) &-0.0187  &-11116 &$V$&INGT  &5295.5787(0.37)&-0.0134  &-9641  &$V$&INGT \\
4536.4510(35)  &-0.0263 &-13013&$V$    &INGT &4964.6351(39) &-0.0262  &-11111 &$V$&INGT  &5297.5974(0.49)&-0.0208  &-9632  &$V$&INGT \\
4539.6050(50)  &-0.0241 &-12999&$V$    &INGT &4965.5461(59) &-0.0157  &-11107 &$V$&INGT  &5298.4989(0.32)&-0.0198  &-9628  &$V$&INGT \\
4555.5957(21)  &-0.0171 &-12928&$V$    &INGT &4966.6605(48) &-0.0269  &-11102 &$V$&INGT  &5304.5789(0.41)&-0.0181  &-9601  &$V$&INGT \\
4558.5209(50)  &-0.0185 &-12915&$V$    &INGT &4967.5715(48) &-0.0164  &-11098 &$V$&INGT  &5307.4995(0.37)&-0.0241  &-9588  &$V$&INGT \\
4573.6128(78)  &-0.0098 &-12848&$V$    &INGT &4968.4678(39) &-0.0205  &-11094 &$V$&INGT  &5308.6352(0.48)&-0.0140  &-9583  &$V$&INGT \\
4591.6039(32)  &-0.0286 &-12768&$V$    &INGT &4969.5905(47) &-0.0235  &-11089 &$V$&INGT  &5309.5267(0.51)&-0.0230  &-9579  &$V$&INGT \\
4593.6353(27)  &-0.0233 &-12759&$V$    &INGT &4970.4921(36) &-0.0224  &-11085 &$V$&INGT  &5318.5369(0.82)&-0.0178  &-9539  &$V$&INGT \\
4596.5626(37)  &-0.0226 &-12746&$V$    &INGT &4971.6100(37) &-0.0301  &-11080 &$V$&INGT  &5319.6648(0.37)&-0.0155  &-9534  &$V$&INGT \\
4597.4642(43)  &-0.0214 &-12742&$V$    &INGT &4972.5159(55) &-0.0247  &-11076 &$V$&INGT  &5321.4552(0.36)&-0.0260  &-9526  &$V$&INGT \\
4598.5876(41)  &-0.0237 &-12737&$V$    &INGT &4973.6431(43) &-0.0232  &-11071 &$V$&INGT  &5322.5825(0.36)&-0.0243  &-9521  &$V$&INGT \\
4608.5046(74)  &-0.0121 &-12693&$V$    &INGT &4974.5450(49) &-0.0217  &-11067 &$V$&INGT  &5324.6152(0.42)&-0.0177  &-9512  &$V$&INGT \\
4609.6179(76)  &-0.0244 &-12688&$V$    &INGT &4975.6699(22) &-0.0224  &-11062 &$V$&INGT  &5331.5982(0.37)&-0.0136  &-9481  &$V$&INGT \\
4611.6472(33)  &-0.0212 &-12679&$V$    &INGT &4976.5682(57) &-0.0247  &-11058 &$V$&INGT  &5333.6231(0.54)&-0.0147  &-9472  &$V$&INGT \\
4613.6671(69)  &-0.0274 &-12670&$V$    &INGT &4977.4694(50) &-0.0239  &-11054 &$V$&INGT  &5334.5176(0.35)&-0.0207  &-9468  &$V$&INGT \\
4614.5766(69)  &-0.0184 &-12666&$V$    &INGT &4978.5984(58) &-0.0205  &-11049 &$V$&INGT  &5335.6492(0.41)&-0.0147  &-9463  &$V$&INGT \\
4615.4670(77)  &-0.0286 &-12662&$V$    &INGT &4979.4993(52) &-0.0201  &-11045 &$V$&INGT  &5336.5411(0.36)&-0.0234  &-9459  &$V$&INGT \\
4619.5244(84)  &-0.0233 &-12644&$V$    &INGT &4980.6185(36) &-0.0265  &-11040 &$V$&INGT  &5337.6671(0.52)&-0.0230  &-9454  &$V$&INGT \\
4620.6515(43)  &-0.0218 &-12639&$V$    &INGT &4981.5197(53) &-0.0258  &-11036 &$V$&INGT  &5343.5241(0.63)&-0.0192  &-9428  &$V$&INGT \\
4896.6416(41)  &-0.0326 &-11413&$V$    &INGT &4982.6499(35) &-0.0213  &-11031 &$V$&INGT  &5348.4781(0.39)&-0.0179  &-9406  &$V$&INGT \\
4910.6146(40)  &-0.0172 &-11351&$V$    &INGT &4983.5402(65) &-0.0314  &-11027 &$V$&INGT  &5350.5047(0.47)&-0.0174  &-9397  &$V$&INGT \\
4912.6332(34)  &-0.0247 &-11342&$V$    &INGT &5240.6413(47) &-0.0208  &-9885  &$V$&INGT  &5617.5019(0.78)&-0.0161  &-8211  &$V$&INGT \\
4913.5382(32)  &-0.0202 &-11338&$V$    &INGT &5249.6515(66) &-0.0155  &-9845  &$V$&INGT  &5618.6192(0.62)&-0.0244  &-8206  &$V$&INGT \\
4919.6161(47)  &-0.0206 &-11311&$V$    &INGT &5265.6296(35) &-0.0212  &-9774  &$V$&INGT  &5619.5208(0.57)&-0.0232  &-8202  &$V$&INGT \\
4921.6366(32)  &-0.0262 &-11302&$V$    &INGT &5267.6546(35) &-0.0223  &-9765  &$V$&INGT  &5620.6495(0.27)&-0.0202  &-8197  &$V$&INGT \\
4923.6640(38)  &-0.0249 &-11293&$V$    &INGT &5268.5603(32) &-0.0170  &-9761  &$V$&INGT  &5621.5492(0.39)&-0.0210  &-8193  &$V$&INGT \\
4924.5722(59)  &-0.0172 &-11289&$V$    &INGT &5269.4593(62) &-0.0185  &-9757  &$V$&INGT  &5622.6737(0.34)&-0.0221  &-8188  &$V$&INGT \\
4925.4686(53)  &-0.0213 &-11285&$V$    &INGT &5270.5812(39) &-0.0223  &-9752  &$V$&INGT  &5623.5817(0.62)&-0.0146  &-8184  &$V$&INGT \\
4926.5957(39)  &-0.0198 &-11280&$V$    &INGT &5271.4820(59) &-0.0219  &-9748  &$V$&INGT  &5646.5382(0.54)&-0.0206  &-8082  &$V$&INGT \\
4927.4840(43)  &-0.0320 &-11276&$V$    &INGT &5272.6103(38) &-0.0193  &-9743  &$V$&INGT  &5647.6642(0.40)&-0.0202  &-8077  &$V$&INGT \\
4935.6006(39)  &-0.0198 &-11240&$V$    &INGT &5273.5063(61) &-0.0238  &-9739  &$V$&INGT  &5648.5741(0.51)&-0.0109  &-8073  &$V$&INGT \\
4936.4974(38)  &-0.0235 &-11236&$V$    &INGT &5275.5328(48) &-0.0234  &-9730  &$V$&INGT  &5649.4581(0.52)&-0.0274  &-8069  &$V$&INGT \\
4938.5190(33)  &-0.0280 &-11227&$V$    &INGT &5276.6617(39) &-0.0200  &-9725  &$V$&INGT  &6736.7630(0.02)&0.0460   &-3239.5&$V$&INGT \\
4939.6506(34)  &-0.0220 &-11222&$V$    &INGT &5277.5637(55) &-0.0185  &-9721  &$V$&INGT  &6736.8722(0.03)&0.0426   &-3239  &$V$&INGT \\
4940.5518(40)  &-0.0213 &-11218&$V$    &INGT &5278.4612(31) &-0.0215  &-9717  &$V$&INGT  &               &         &       &   & \\
\hline\hline
\end{tabular}

\begin{list}{}{  }
\item[Ref:]{INGT means Isaac Newton Group of Telescopes, Apartado de Correos 321, E-38700
Santa Cruz de La Palma, Tenerife, Spain.}
\end{list}
\end{table}

\section{Orbital Period Investigation}
The $(O-C)$ method is the traditional way to reveal variations on orbital period.
Before the present work, only one minima time of J1405 had been published.
For analysing the period changes of the system, we collected all the CCD times of light minimum.
Mr. Marcus Lohr sent us 121 minima times from Super WASP, and the LCs obtained from OAN offered another 4.
Minimum times with the same epoch have been averaged, and only the mean values are listed in Table 3.
In our fitting process, according to determined errors listed in Table 3, the weight of the Super WASP data is 1 and that of our data is 5.

The $(O-C)$ diagram shows a upward parabola variation and the fitting curve is plotted in Figure 2.
Based on the least-square method, the new ephemeris
\begin{eqnarray}
{\rm Min.~I}&=&2457470.95618(\pm0.0003) \nonumber\\
    &  & +0.^{d}22512639(\pm0.00000003)\times{E}\nonumber\\
    & &+1.29(\pm0.02)\times{10^{-10}}\times{E^{2}}\nonumber\\
\end{eqnarray}
was obtained. With the quadratic term included in this equation, a
secular period increase rate is determined: dP/dt=2.09$\times$$10^{-7}$days yr$^{-1}$.
\begin{figure}{!h}
\begin{center}
\includegraphics[width=11cm]{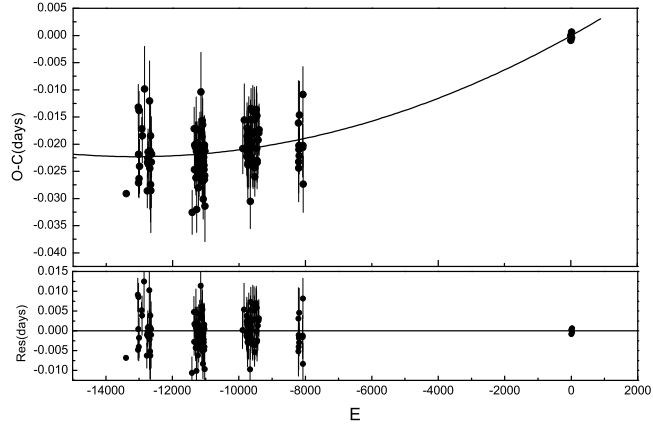}
\caption{ The $(O - C)$ diagram of J1405 formed by all available
measurements. The $(O - C)$ values were computed by using a newly
determined linear ephemeris (Eq.1). The solid line represents the quadratic fit (Eq.2). The bottom
panel plots the residuals for Equation 2.}
\end{center}
\end{figure}
\section{Photometric Solutions}
To derive its physical parameters, the 2013 version of Wilson-Devinney (W-D) program (Wilson 1971; Wilson \& Van 2003; Wilson et al. 2010) is used.
The number of observational data applied in the program are 81 in $B$ band, 78 in $V$ band, and 72 in $R$ band.

Before analyzing the LCs, the value of some input parameter were set.
The temperature for star 1 (star eclipsed at the primary light minimum), $ T_{1}=4680K$ was fixed according to mean color index (Cox 2000).
We take the same values of gravity-darkening coefficients and the bolometric albedo for both components, i.e., $g_{1}=g_{2}=0.32$
according to the stellar temperatures given by Claret (2000) and $A_{1}=A_{2}=0.5$ (Lu \& Rucinski 1993) were set for late-type
stars with a convective envelope.
The bolometric and bandpass limb-darkening coefficients were chosen from Van Hamme (1993).
To account for the limb-darkening in detail, logarithmic functions are used.
These fixed parameters are listed in Table 4.
The adjustable parameters were: the mass ratio, $q$; the orbital inclination, $i$; the mean temperature of star 2, $T_{2}$;
the dimensionless potentials of the two components $\Omega_{1}$ and $\Omega_{2}$;
the monochromatic light of star 1, $L_{1B}$, $L_{1V}$, and $L_{1R}$.

Two sets of LCs are obtained, but, the data from March 17 are of better quality.
So, we started the analysis with this set of LCs.
Mode 3 for contact binary system is adopted ($\Omega_{1}=\Omega_{2}$ in this case).
Because there is no spectroscopic observations for J1405 published, we used a q-search method (fix $q$) to obtain initial input parameters.
The solutions are carried out with mass ratios ranging from less than 0.25 to larger than 3.0.
The relation between the sum of weighted square deviations $\Sigma\omega(O - C)^{2}$ and $q$ is plotted in Figure 3.
The $\Sigma - q$ curves shown a lowest value at $q=1.55$.
We then set the initial value of $q$ to be 1.55 and treated it as an adjustable parameter.
After all the free parameters converged, one set of solutions was derived.
Because our LCs are asymmetric, and the main reason is the variation of spot location and size(Kang et al. 2002).
Considering this case, cool star-spots model was used to get better solutions.
As we all know, four parameters to describe a spot, namely the temperature factor, $T_{s}$;
the latitude, $Lat$; the longitude $Lon$ and the angular radius $Rad$.
During our analysis, only the $T_{s}$ was fixed at a series of trial values until we found the best solution.
Finally, we found that one cool star-spot on the secondary component could get the best fit,
and the fitting residual of our solution with spot is much smaller than that without spot.
Therefore, we adopted the cool star-spot model as the final solution.
The derived photometric solutions are listed in the Table 5 and the theoretical LCs computed with the cool star-spot model are plotted in the Figure 4.
Besides, we also get the geometrical structure of the system, which is shown in Figure 5.
Because the LCs obtained on March 22 are of worse quality, and we didn't analyze them.

\begin{figure}
\begin{center}
\includegraphics[width=8cm,height=6cm]{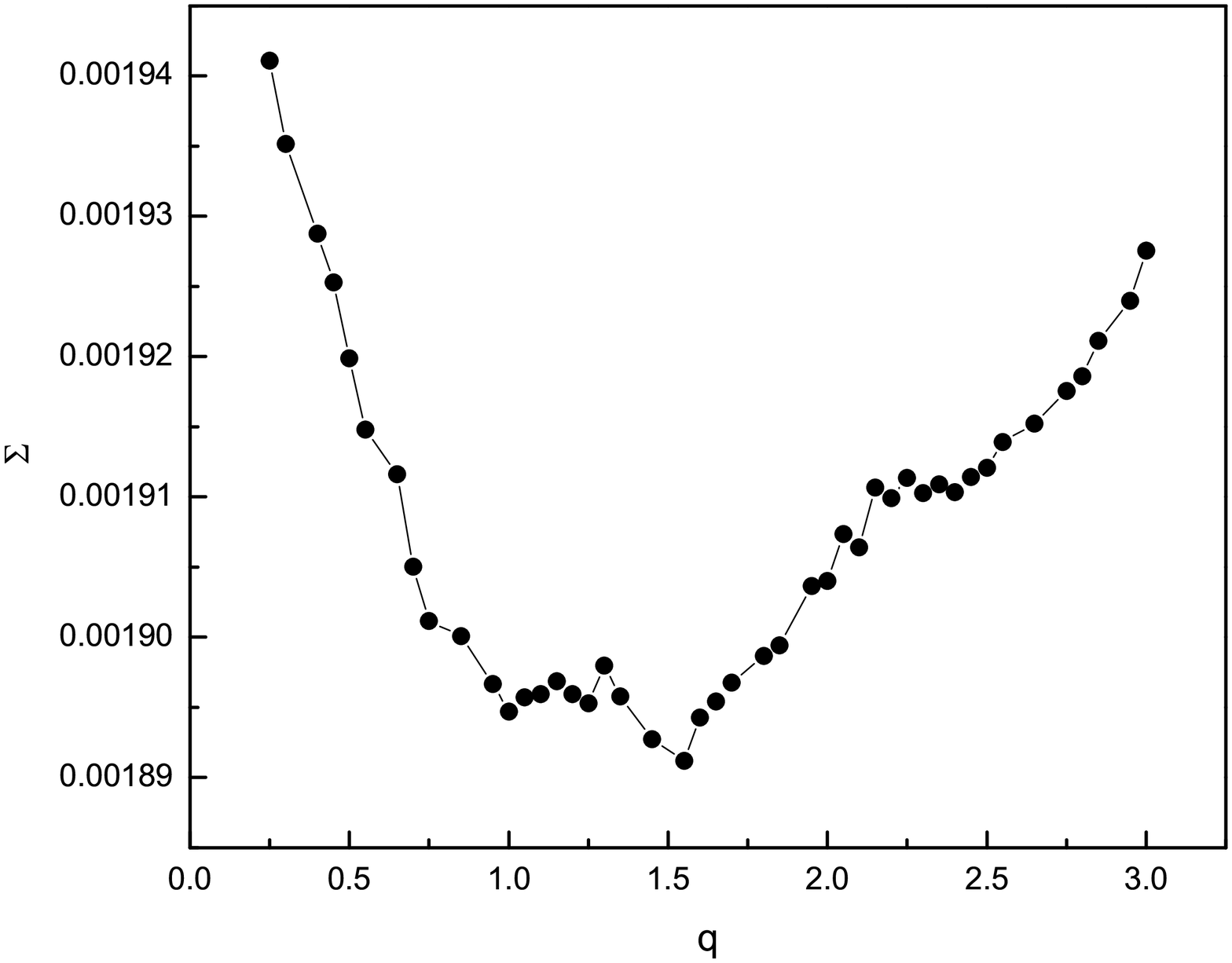}
\caption{The $\Sigma - q$ curve for J1405. The minimum residuals is at $q=1.55$.}
\end{center}
\end{figure}

\begin{figure}
\begin{center}
\includegraphics[width=10cm,height=8cm]{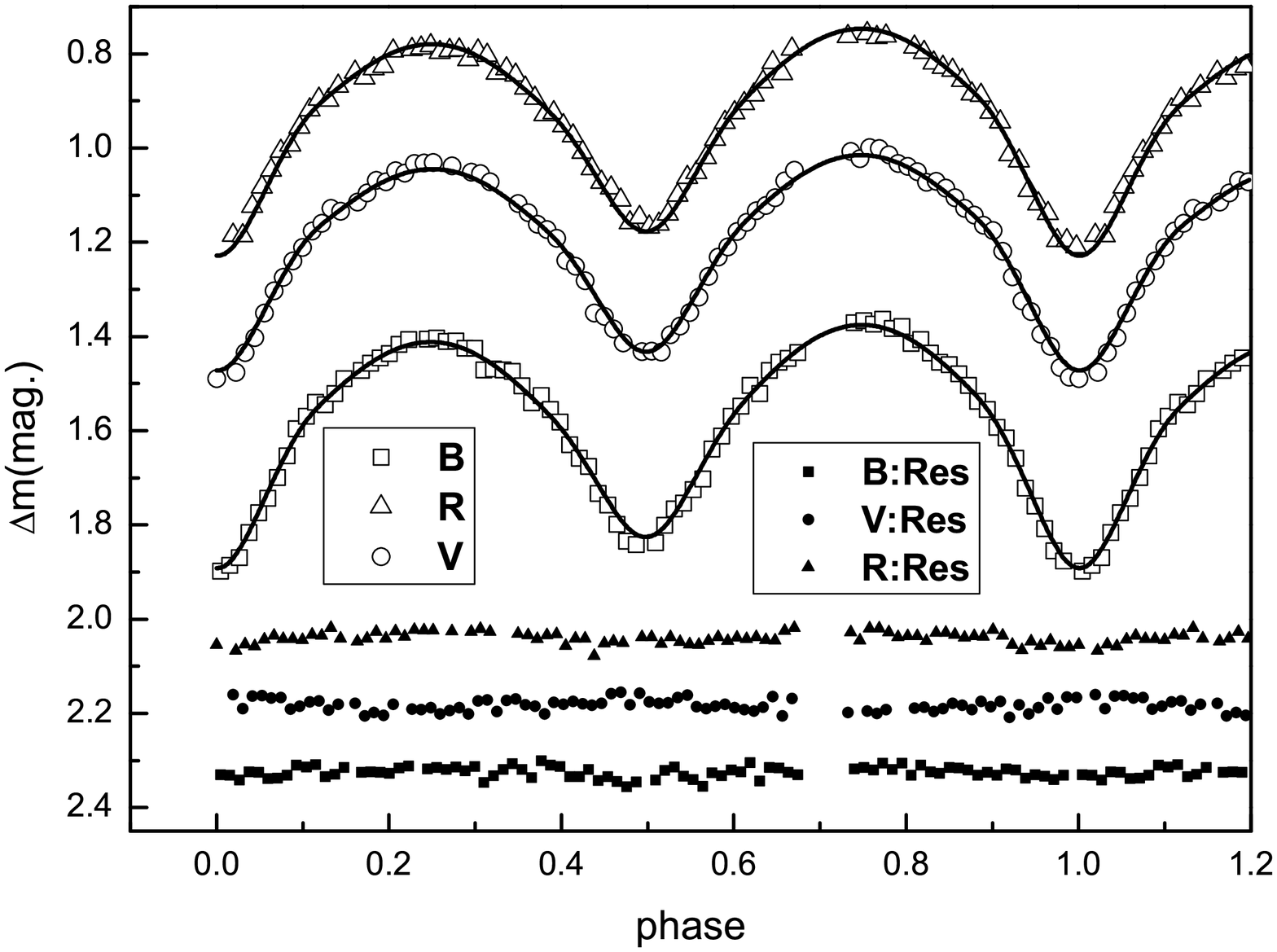}
\caption{Observed (open cycle, triangle and square) and theoretical light curves (solid line) calculated with cool
star-spot listed in Table 5. Residuals from the solutions are shown in the bottom panel.}
\end{center}
\end{figure}

\begin{figure}
\begin{center}
\includegraphics[width=12cm]{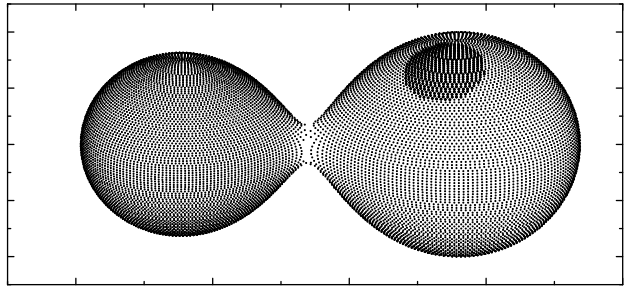}
\caption{Geometric structure of J1405 at phase 0.25.}
\end{center}
\end{figure}

\begin{table}[!h]
\caption{Fixed Parameters During Photometric Analysis}
\begin{center}
\small
\begin{tabular}{lcc}
\hline\hline
Parameters              & $\qquad$$\qquad$$\qquad$$\qquad$$\qquad$$\qquad$$\qquad$$\qquad$     & Values\\
\hline
$g_1=g_2$               & $\qquad$$\qquad$$\qquad$$\qquad$$\qquad$$\qquad$$\qquad$$\qquad$     & 0.32          \\
$A_1=A_2$               & $\qquad$$\qquad$$\qquad$$\qquad$$\qquad$$\qquad$$\qquad$$\qquad$     & 0.50          \\
$x_{1bolo}$,$x_{2bolo}$ & $\qquad$$\qquad$$\qquad$$\qquad$$\qquad$$\qquad$$\qquad$$\qquad$     & 0.641, 0.638  \\
$y_{1bolo}$,$y_{2bolo}$ & $\qquad$$\qquad$$\qquad$$\qquad$$\qquad$$\qquad$$\qquad$$\qquad$     & 0.172, 0.163  \\
$x_{1B}$,$x_{2B}$       & $\qquad$$\qquad$$\qquad$$\qquad$$\qquad$$\qquad$$\qquad$$\qquad$     & 0.848, 0.844  \\
$y_{1B}$,$y_{2B}$       & $\qquad$$\qquad$$\qquad$$\qquad$$\qquad$$\qquad$$\qquad$$\qquad$     & 0.096, 0.129 \\
$x_{1V}$,$x_{2V}$       & $\qquad$$\qquad$$\qquad$$\qquad$$\qquad$$\qquad$$\qquad$$\qquad$     & 0.802, 0.801  \\
$y_{1V}$,$y_{2V}$       & $\qquad$$\qquad$$\qquad$$\qquad$$\qquad$$\qquad$$\qquad$$\qquad$     & 0.045, 0.022  \\
$x_{1R},x_{2R}$         & $\qquad$$\qquad$$\qquad$$\qquad$$\qquad$$\qquad$$\qquad$$\qquad$     & 0.749, 0.755  \\
$y_{1R}$,$y_{2R}$       & $\qquad$$\qquad$$\qquad$$\qquad$$\qquad$$\qquad$$\qquad$$\qquad$     & 0.123, 0.108  \\
\hline
\end{tabular}
\end{center}
\end{table}

\begin{table}
\begin{center}
\caption{Photometric Solutions for J1405}
\begin{tabular}{lllll}
\hline\hline
Parameters                   &   March 17th   && March 17th&\\
                             &   No spot      &Errors& With spot &Errors\\
\hline
$T_1$(K)                     &4680             &Assumed        &4680    & Assumed          \\
$q$                          &1.5501           &$\pm$ 0.0199   &1.5488  & $\pm$ 0.0163     \\
$T_2$(K)                     &4563             &$\pm$ 16       &4523    & $\pm$ 21         \\
$i$($^{\circ}$)                 &68.996           &$\pm$ 0.257    &68.616  & $\pm$ 0.321      \\
$L_1/(L_1+L_2)(B)$           &0.4545           &$\pm$ 0.0074   &0.4742  & $\pm$ 0.0135     \\
$L_1/(L_1+L_2)(V)$           &0.4434           &$\pm$ 0.0058   &0.4590  & $\pm$ 0.0121     \\
$L_1/(L_1+L_2)(R)$           &0.4342           &$\pm$ 0.0045   &0.4466  & $\pm$ 0.0112     \\
$\Omega_1$=$\Omega_2$        &4.5612           &$\pm$ 0.0092   &4.5528  & $\pm$ 0.0026     \\
$r_1(pole)$                  &0.3237           &$\pm$ 0.0009   &0.3247  & $\pm$ 0.0011     \\
$r_1(side)$                  &0.3391           &$\pm$ 0.0011   &0.3402  & $\pm$ 0.0015     \\
$r_1(back)$                  &0.3733           &$\pm$ 0.0016   &0.3751  & $\pm$ 0.0022     \\
$r_2(pole)$                  &0.3967           &$\pm$ 0.0009   &0.3976  & $\pm$ 0.0014     \\
$r_2(side)$                  &0.4201           &$\pm$ 0.0011   &0.4213  & $\pm$ 0.0018     \\
$r_2(back)$                  &0.4510           &$\pm$ 0.0015   &0.4526  & $\pm$ 0.0025     \\
f(\%)                        &6.7              &$\pm$ 1.6      &7.9     & $\pm$ 0.5        \\
Latitude($^{\circ}$)         &                 &               &334.392 & $\pm$ 4.242     \\
Longitude($^{\circ}$)        &                 &               &253.902 & $\pm$ 6.187     \\
Radius(radian)               &                 &               &0.356   & $\pm$ 0.089     \\
$T_{s}$                      &                 &               &0.85    &  Assumed        \\
$\sum{\omega_{i}(O-C)_i^2}$  & 0.001891        &               &0.001135&                 \\
\hline
\end{tabular}
\end{center}
\end{table}

\section{Discussions and Conclusions}
The asymmetry displayed by the observed LCs suggests a spot activity of the system.
Because of that, we analyzed the LCs and obtained the photometric solutions with a cool star-spots on the more massive component using the 2013 version of the Wilson-Devinney code.
The results suggest that J1405 is a W-subtype contact eclipsing binary near the short-period cutoff with a mass ratio of $q$ = 1.55 $\pm$0.02.
The mean contact degree ($f = 7.3\,\%$) reveals that it is a shallow contact system with similar surface temperature of the components ($\Delta T = 140K$).
Similar contact binaries including AH Vir (Lu \& Rucinski 1993; Kjurkchieva et al. 2015), RZ Com (He \& Qian 2008; Xiang \& Zhou 2004),
AM Leo (Hille et al. 2004), U Peg (Mohajerani \& Percy 2011; Djura$\breve{s}$evi$\acute{c}$ et al. 2001) and SW Lac ($\c{S}$enavc${\i}$ 2012).
The W Uma-type binaries are formed from initially detached binaries by angular momentum
loss (AML) via magnetic braking (Qian et al. 2013).
As other late K-type contact binaries with the short-period, J1405 is also in marginal contact and presents remarkable
asymmetric light curves, representing probable surface activities (Zhang et al. 2014; Jiang et al. 2015a).
Just as Qian et al. (2015a) discussed, the orbital shrinkage due to AML may result in the formation of a contact system similar to J1405.
The progenitor of J1405 may be a short-period detached EB system similar to DV Psc (Pi et al. 2014),
and the J1405 may be at the same evolutional phase like the 1SWASP J015100.23-100524.2 (Qian et al. 2015a).

Another feature of J1405 is its spot activity.
Generally, for late-type contact binary stars, their deep convective envelope
along with fast rotation can help to produce a strong magnetic dynamo,
because of that, they will display some solar-like activity such as photospheric cool star-spots (Li et al. 2015).
A cool star-spot is known as the strong magnetic area which can change the shape of LCs.
We adopted cool star-spot model of the (W-D) program with one spot on the secondary component to explain it.
Just as the Figure 4 shown, the fitting LCs with cool star-spot coincide very well with the observational data at all phases.
Therefore, using cool star-spot model to explain the asymmetry of LCs are reasonable.

Based on the analysis of the $(O-C)$ diagram, we found that the orbital period of
J1405 show a upward parabolic variation, which represents an increase of the period.
According to the obtained parameters, the rate of dP/dt=2.09$\times$$10^{-7}$days yr$^{-1}$ is derived.
The secular increase of the orbital period may be interpreted as the mass transfer from the less massive component to the more massive one,
such as EP And (Liao et al. 2013), 1SWASP J074658.62+224448.5 (Jiang et al. 2015a), 1SWASP J075102.16+342405.3 (Jiang et al. 2015b).
However, the time span of our data is only 10 years, the increase of the period might be only a part of long-term changes.
Further observations are required to confirm this result.

\begin{acknowledgements}
This work is partly supported by Chinese Natural Science Foundation (No.11133007, 11573063),
the Science Foundation of Yunnan Province (grant No. 2012HC011).
Many thanks to Dr.Marcus Lohr for his kindly sending us eclipse timings of J1405.
We are also especially indebted to the anonymous referees who given us
useful comments and cordial suggestions, which helped us to improve
the paper greatly.
\end{acknowledgements}

\label{lastpage}
\end{document}